\documentclass[10pt]{ismd08}
\usepackage{wrapfig}
\usepackage{graphicx}
\usepackage{cite,./mcite}

\usepackage{amsmath}
\usepackage{amssymb}

\def\met{\mathbin{E\mkern - 11mu/_T}}

\begin{document}
\title{Diffractive Production of Jets and
  Vector Bosons at the Tevatron}
\author{Kenichi Hatakeyama\protect\footnote{\ \ speaker},
 for the CDF Collaboration}
\institute{$^1$Rockefeller University}
\maketitle
\begin{abstract}
Recent results on diffractive dijet and vector boson production and
exclusive dijet production from the Collider Detector at Fermilab
(CDF) experiment are presented.
\end{abstract}

\section{Introduction}

CDF Collaboration performed various measurements on inclusive
diffraction and exclusive production using $p\bar p$ collision data
from the Fermilab Tevatron collider collected in Run I (1992--1996)
and Run II (2001--).
One of the important results from the Run I studies is the observation
of the QCD factorization breakdown in hard single diffractive (SD)
processes~\cite{DiffJJ-CDF,DiffW-CDF,Diffb-CDF,RPJJ-CDF,RPJJ-630-CDF};
the rate of hard SD processes, in which one of the
incoming proton or antiproton is scattered quasielastically and a hard
partonic scattering (such as dijet production) occurs, was found
to be lower than theoretical predictions by a factor of ${\cal
  O}(10)$.
In \cite{RPJJ-CDF,RPJJ-630-CDF}, the diffractive structure function
$F^D(Q^2,x,\xi,t)$ was measured using SD dijet
events and found to be suppressed with respect to the one measured
in $ep$ collisions at HERA by ${\cal O}(10)$, where $\xi$ is the
fractional momentum loss of the diffracted (anti)proton and
$t$ is the four-momentum transfer squared.
This suppression is similar to the one observed in soft diffractive
processes with respect to the Regge theory predictions,
and is generally attributed to additional color exchanges in the same
$p\bar p$ collision which spoil the diffractive rapidity
gap~\cite{Gotsman:1999xq,Kaidalov:2001iz,Dino}.

Another important result from the Run I diffractive studies is from
a study on $F^D$ using double pomeron exchange (DPE) dijet
events~\cite{DPEJJ},
$p+\bar p \to p + jjX + \bar p$.
The diffractive structure function $F^D$ measured in DPE dijet events
was found to be approximately equal to expectations from HERA.
This observation is consistent with the expectations from, {\it e.g.},
the gap probability renormalization model~\cite{Dino}.
The main goal of the Run II diffractive studies is to study
the characteristics of diffractive events more in detail with help of
the upgraded detectors and larger statistics in order to deepen our
understanding of diffractive exchange and the QCD nature of the
pomeron.

\begin{figure}[thb]\centering\leavevmode
  \includegraphics[width=0.25\hsize,clip=]
  {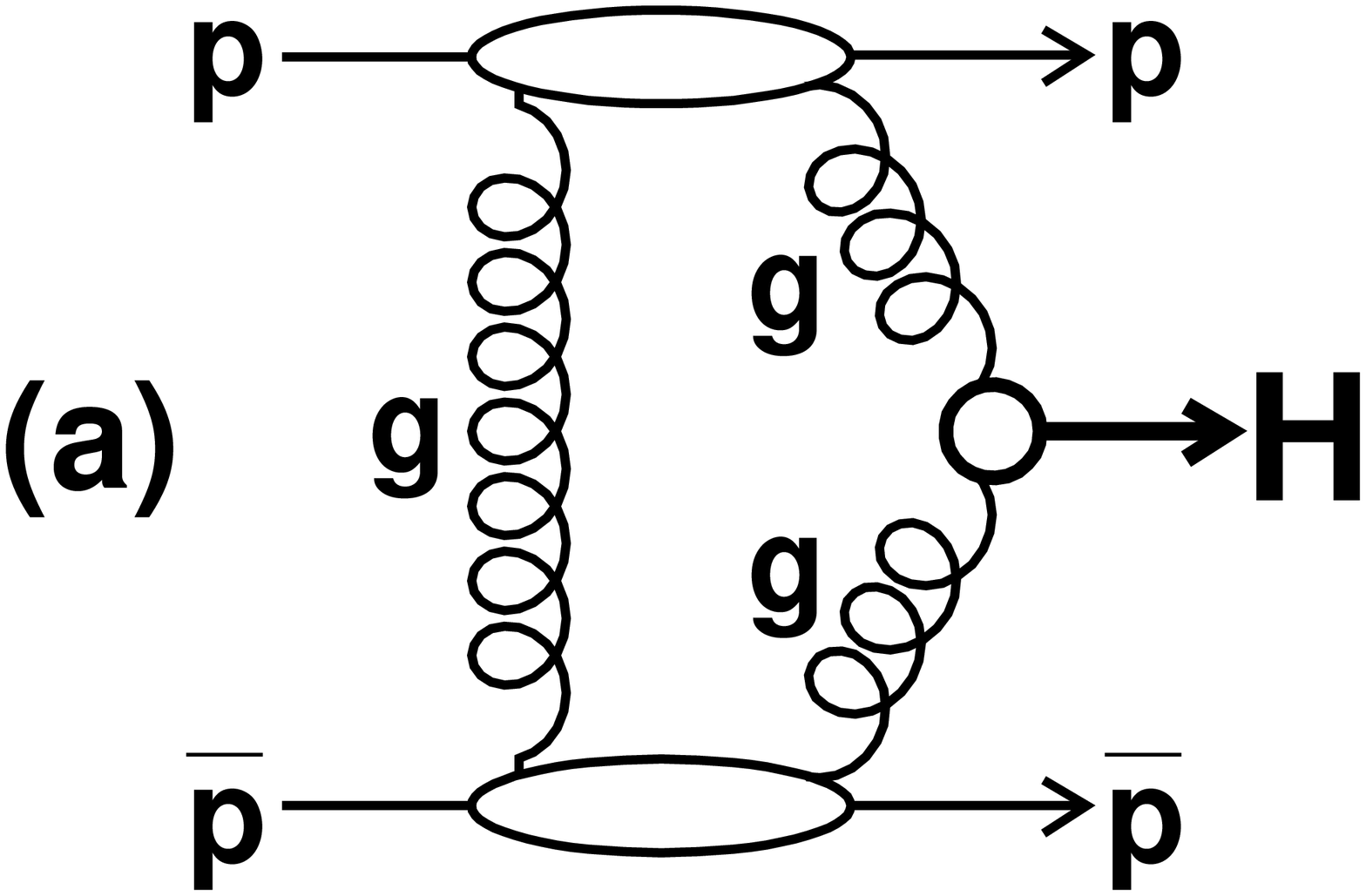}
  \hspace*{1cm}
  \includegraphics[width=0.25\hsize,clip=]
  {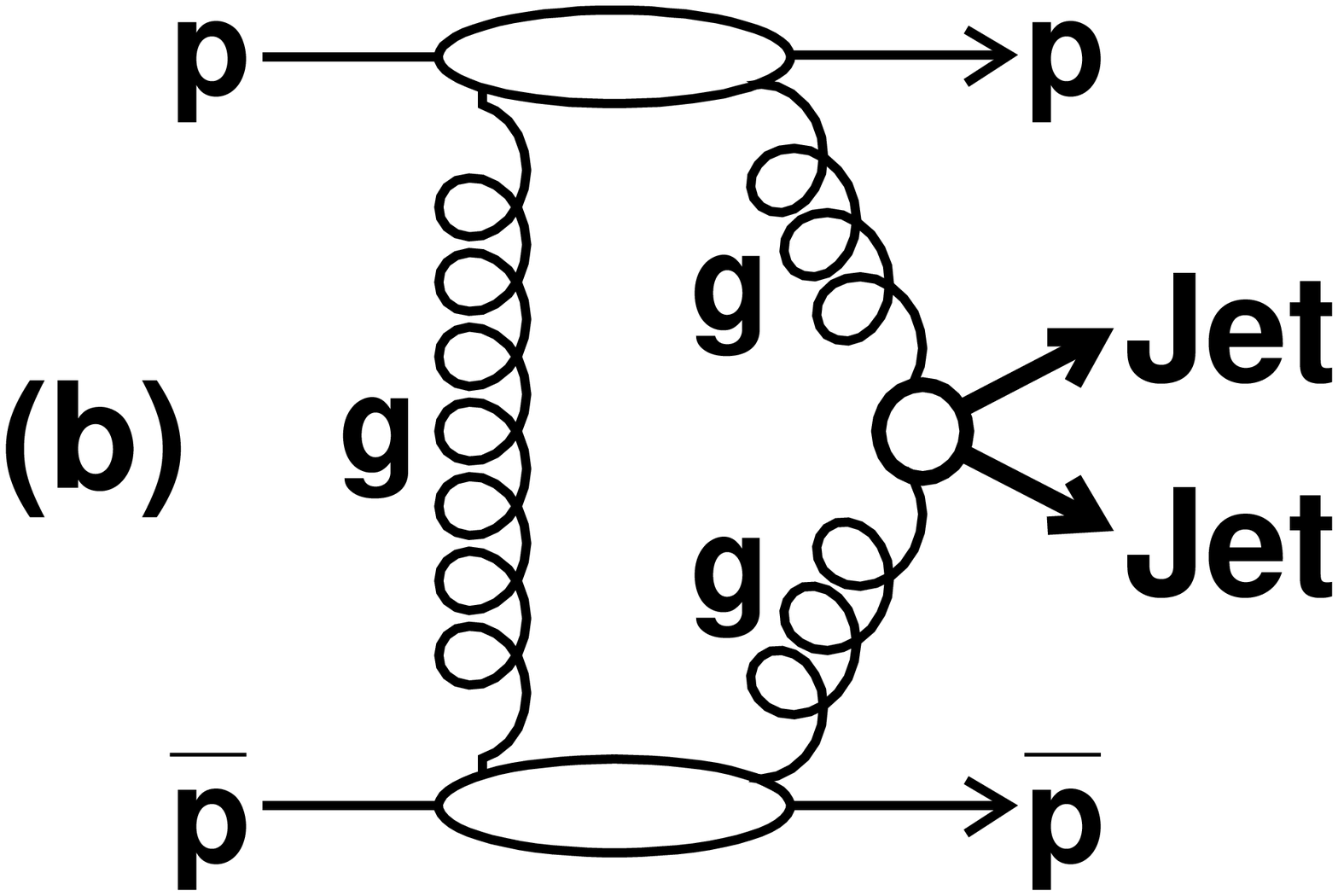}
  \caption{Diagrams for exclusive production of (left) Higgs and (right)
    dijet production.}
  \label{fig:excl_jj_feynman}
\end{figure}

In addition, there has been an increased interest in studies on 
{\it exclusive} events, mainly due to a possibility of finding the
Higgs boson in exclusive events at the Large Hadron Collider (LHC).
Exclusive events in $pp$ ($p\bar p$) collisions contain nothing but
the leading proton and (anti)proton and the object(s) of interest
such as dijet, diphoton, dielectron, and most importantly the Higgs
boson, as shown in Fig.~\ref{fig:excl_jj_feynman}.
We do not expect to observe the exclusive Higgs production at the
Tevatron; however, we can study other exclusive processes that can
provide a calibration for theoretical predictions of exclusive Higgs
production at the LHC.

The recent Run II studies on hard diffraction and exclusive dijet
production are presented below.
Studies on exclusive dilepton, diphoton, and charmonium states are
presented in~\cite{Mike}.

\section{Diffractive Dijet Production}

The diffractive structure function $F^D$ was studied using 
Run II SD dijet data using a similar way to that used
in Run I studies~\cite{RPJJ-CDF,RPJJ-630-CDF}, {\it i.e.}, by taking a
ratio of SD to non-diffractive (ND) dijet rates as a
function of $x$, which is in leading-order QCD approximately equal to
the ratio of diffractive to ND structure function.

One of the major challenges in Run II diffractive studies is the
rejection of multiple $p\bar p$ interaction events, in which
diffractive rapidity gaps are spoiled by overlapping $p\bar p$
interactions (overlaps) and the hard scattering cannot be associated with the
diffracted leading (anti)proton accurately.
This rejection was done by reconstructing $\xi$ from the calorimeter
towers by $\xi^{cal}=\sum_{\rm towers} E_T^{i}\eta^i / \sqrt{s}$;
$\xi^{cal}\sim\xi^{RP} < 0.1$ in SD events without
overlaps, while $\xi^{cal}>0.1$ in events with overlaps,
where $\xi^{RP}$ refers to the $\xi$ value
reconstructed based on the information from the Roman pot (RP) detector
which detects the diffracted antiproton.

The high statistics Run II data allowed the SD/ND dijet ratio
measurement in $Q^2$ up to $10^4$ GeV$^2$, and no appreciable $Q^2$
dependence was observed.
Also in the Run II study, the $t$ distribution in SD dijet
events was measured up to $Q^2\sim 4500$ GeV$^2$, and no dependence of
the shape of the $t$ distribution on $Q^2$ was found.

\section{Diffractive W/Z Production}

CDF studied diffractive $W/Z$ production using the Run II data recently.
The study of diffractive $W/Z$ production is important to determine
the quark content of the pomeron;
the production by gluons is suppressed by a factor of $\alpha_s$
and it can also be identified by an additional jet.

In Run I, CDF studied diffractive $W$ production
by identifying diffractive events using rapidity
gaps~\cite{DiffW-CDF}, and found the fraction of
$W$ events which are diffractive to be 
$[1.15\pm0.51({\rm stat})\pm0.20({\rm syst})]$\%.
In addition, the gluon content of the pomeron was determined
to be $[54^{+16}_{-14}]$\%
in combination with results on diffractive dijet and $b$-quark
production~\cite{Diffb-CDF,DiffJJ-CDF}.
D0 made measurements on diffractive $W$ production and also $Z$
production~\cite{DiffWZ-D0}, and reported
the fractions of $W$ and $Z$ events with a rapidity gap to be
$[0.89^{+0.19}_{-0.17}]$\% and
$[1.44^{+0.61}_{-0.52}]$\%~\cite{DiffWZ-D0}, respectively.
These fractions are not corrected for the gap acceptance correction
$A_{gap}$, {\it i.e.}, the fraction for diffractive events that satisfy the
experimental definitions of the rapidity gaps.
The estimate on $A_{gap}$ ranges from 0.2 to 1.0 depending on the
diffractive models considered.

%\begin{wrapfigure}{r}{0.39\columnwidth}
%\centerline{
%\includegraphics[width=0.39\columnwidth,clip=]{plots/mw_bless.eps}
%}
%\caption{Reconstructed $W$ mass in diffractive $W$ candidate
%  events.}\label{fig:mw}
%\end{wrapfigure}

\begin{figure}[htb]\centering\leavevmode
\includegraphics[width=0.31\columnwidth,clip=]{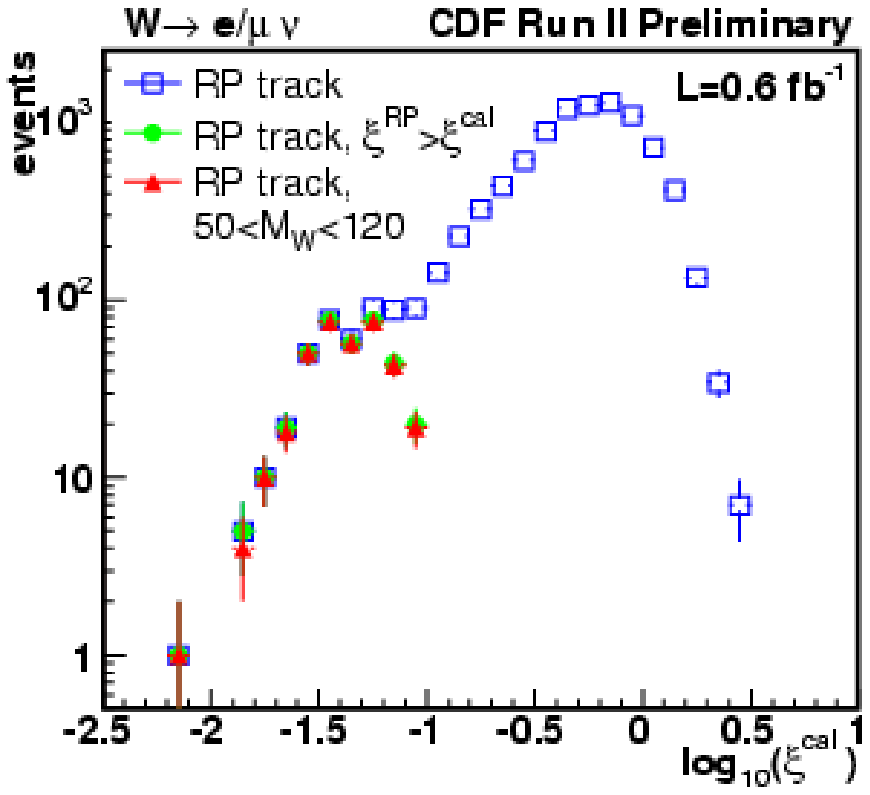}
\includegraphics[width=0.32\columnwidth,clip=]{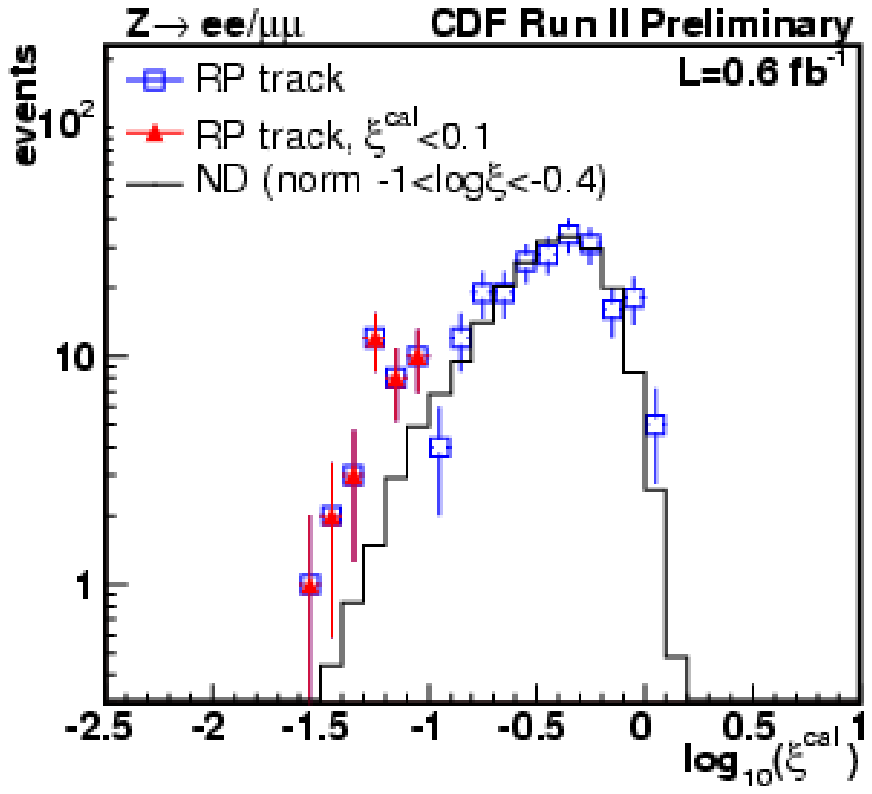}
\includegraphics[width=0.35\columnwidth,clip=]{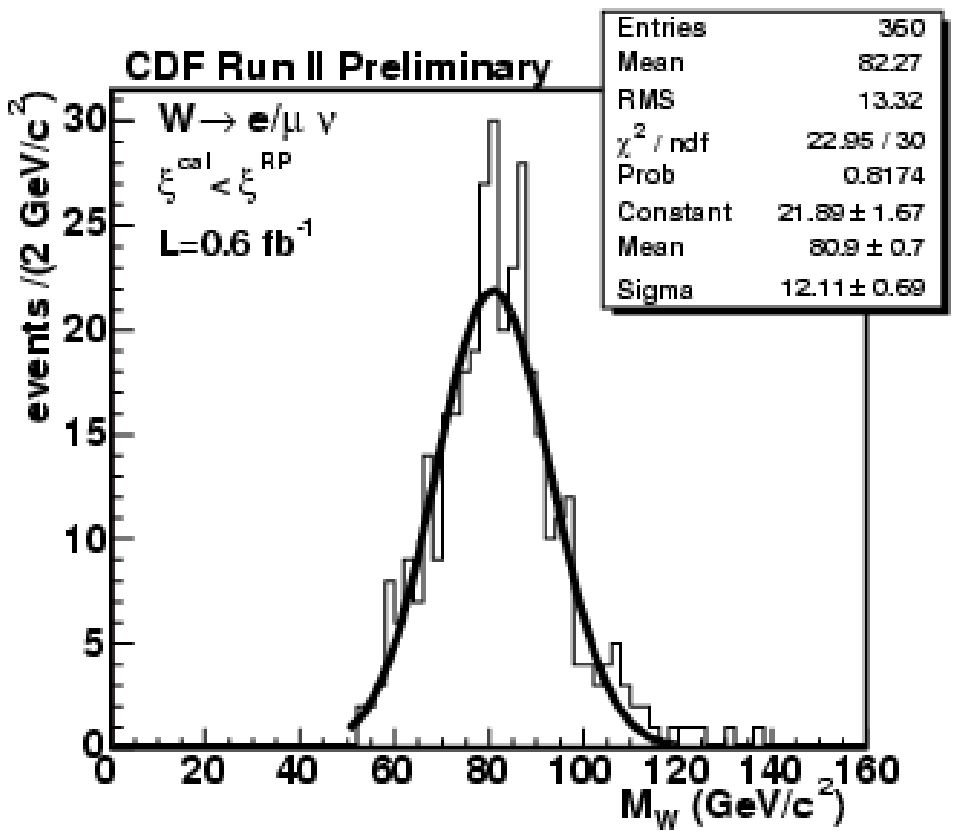}
\caption{Calorimeter distribution in $W$ (left) and $Z$ (center) events
  with a Roman-pot track. (right) Reconstructed $W$ mass in
  diffractive $W$ candidate events.}
\label{fig:xical_wz}
\end{figure}

In the new CDF Run II measurement, the RP detector is
used to detect leading antiproton in diffractive $W/Z$ events.
The RP detector provides an accurate $\xi$ measurement, and also
eliminates the ambiguity associated with $A_{gap}$.
As in diffractive dijet production, $\xi$ can be reconstructed from
both the energy depositions in the calorimeters ($\xi^{cal}$) and hits
in the RP detector ($\xi^{RP}$).
The $\xi^{cal}$ distributions in $W/Z$ events with a leading
antiproton are shown in Fig.~\ref{fig:xical_wz}.
The diffractive $W$ and $Z$ candidate events without overlaps are
selected by requiring $\xi^{cal}<\xi^{RP}$ and $\xi^{cal}<0.1$,
respectively.

In diffractive $W\to l\nu$ events without overlaps, the
difference between $\xi^{cal}$ and $\xi^{RP}$
is related to missing $E_T$ ($\met$) and
$\eta_{\nu}$ as $\xi^{RP} - \xi^{cal} =
\frac{\met}{\sqrt{s}}e^{-\eta_{\nu}}$,
which allows to determine the neutrino kinematics, and consequently
the $W$ kinematics. 
The reconstructed $W$ mass is shown in Fig.~\ref{fig:xical_wz}.

%The distributions of $\xi^{cal}$ in $W$ and $Z$ events with the
%leading antiproton in the Roman pot detector are shown in
%Fig.~\ref{fig:xical_wz}.
%In order to remove events with multiple proton-antiproton
%interactions, in $W$ events, the requirement of $\xi^{cal}<\xi^{RR}$
%was imposed.
%%After correcting for the residual multiple interaction events,
The fractions of $W$ and $Z$ events which are diffractive are measured
to be
$[0.97 \pm0.05({\rm stat}) \pm 0.11({\rm syst})]$\%
and 
$[0.85 \pm0.20({\rm stat}) \pm 0.11({\rm syst})]$\%
in $0.03<\xi<0.10$ and $|t|<1$ GeV/$c$.
The measured diffractive $W$ fraction is consistent with
the Run I CDF result when corrected to the $\xi$ and $t$ range in this
measurement.
%For diffractive $Z$, the candidate events were selected by requiring
%$\xi^{cal}<0.1$ to remove multiple interaction events, and the
%fraction of $Z$ events which are diffractive was measured to be
%$[0.85 \pm0.20({\rm stat}) \pm 0.11({\rm syst})]$ in the same
%kinematic region as in the $W$ measurement.

\section{Rapidity Gaps between Very Forward Jets}

\begin{figure}[htb]\centering\leavevmode
  \includegraphics[width=0.50\hsize,bb=0 0 550 407,clip=]
  {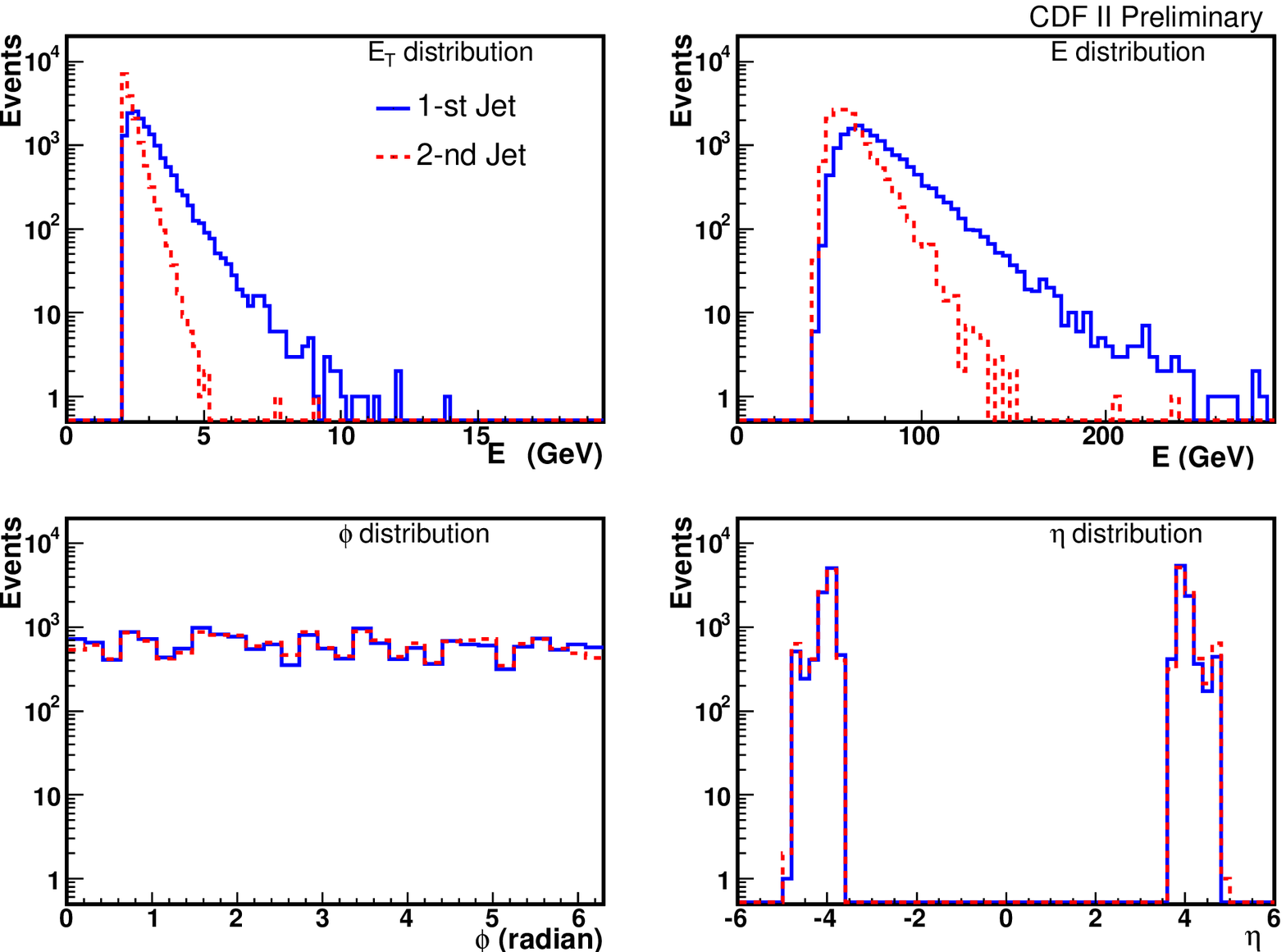}
  \includegraphics[width=0.49\hsize,bb=0 0 520 384,clip=]
  {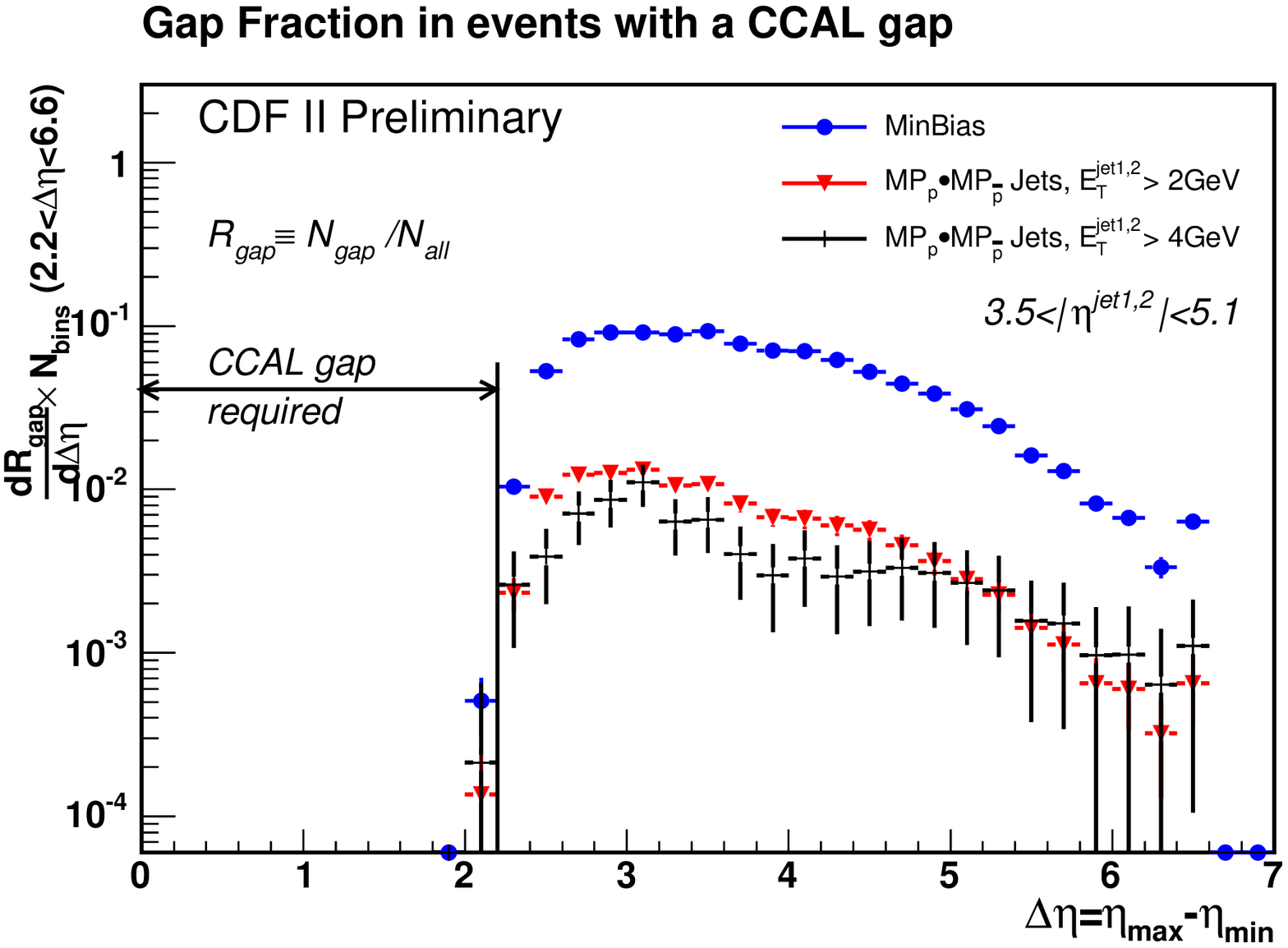}
  \caption{(left) Kinematic distributions for the two leading
    jets in an event, both in the MP calorimeters.
    (right) The gap fraction $R_{gap}$
    vs. $\Delta\eta$
    for min-bias and MP dijet events of $E_T^{jet1,2}>2$ GeV
    and $E_T^{jet1,2}>4$ GeV.}
  \label{fig:Rgap_DeltaEta}
\end{figure}

The double diffractive (DD) dissociation refers to a
class events in which two colliding particles dissociate into clusters
of particles (including jets in the case of hard DD events)
with a large rapidity gap between them.
%Such events are considered to occur due to the exchange of an object
%with vacuum quantum numbers.
%
Measurements on DD events were made by
CDF~\cite{DD-CDF1,DD-CDF2,DD-CDF3,DD-CDF4} and D0~\cite{DD-D0}
in Run I in $p\bar p$ collisions at $\sqrt{s}=1800$ and $630$ GeV.
Recently, CDF reported new preliminary results on events with a
rapidity gap between forward jets from the Run II data.
In CDF II, the miniplug (MP) calorimeters covering
$3.5\lesssim|\eta|\lesssim 5.1$ allow a study of 
very forward jets with a larger rapidity gap between them than
in Run I.
Figure~\ref{fig:Rgap_DeltaEta} (left) shows the kinematic
characteristics of the leading two jets in an event both in MPs
with $E_T>2$ GeV and $\eta_1\eta_2<0$.
Since these jets are in a very forward region, they have high energies
despite their relatively low $E_T$'s.

The dependence of the gap fraction $R_{gap} = N_{gap}/N_{all}$ 
was studied as a function of $\Delta\eta = \eta_{max} - \eta_{min}$
in these MP dijet events in a similar way as in~\cite{DD-CDF4}.
$\eta_{max(min)}$ is the pseudorapidity of the tower closet to
$\eta=0$ in the proton(antiproton) outgoing direction.
The comparison of $R_{gap}$ as a function of
$\Delta\eta$ in min-bias events and MP dijet events with
$E_T^{jet1,2}>2$ GeV and $E_T^{jet1,2}>4$ GeV is shown in
Fig.~\ref{fig:Rgap_DeltaEta}.
A rapidity gap in $|\eta|<1.1$ (CCAL gap) is always required.
The event fraction with a central rapidity gap is about 10\%
in soft events, while it is about 1\% in dijet events,
which is consistent with the results from
Run I~\cite{DD-CDF1,DD-CDF2,DD-CDF3,DD-CDF4,DD-D0}.
It is interesting to note that the shape of the $R_{gap}$
distribution is similar between the soft and hard events.
A study on the azimuthal decorrelation between the two
leading jets in these forward dijet events is underway
in order to investigate the effect of the Muller-Navelet
jets~\cite{Marquet:2007xx}.

\section{Exclusive Dijet Production}

\begin{figure}[htb]\centering\leavevmode
  \includegraphics[width=0.49\hsize,clip=]
  {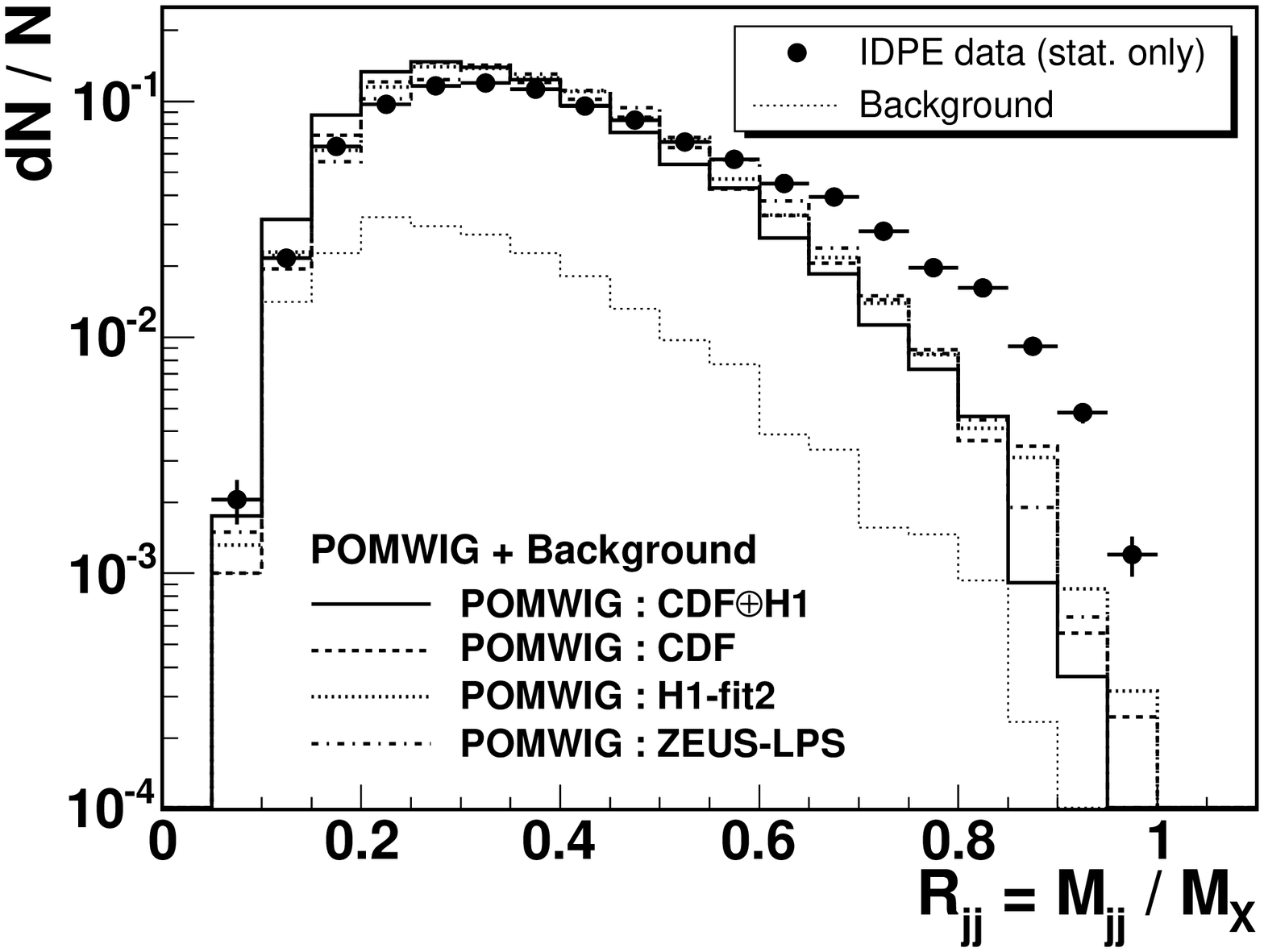}
  \includegraphics[width=0.49\hsize,bb=0 0 567 360,clip=]
  {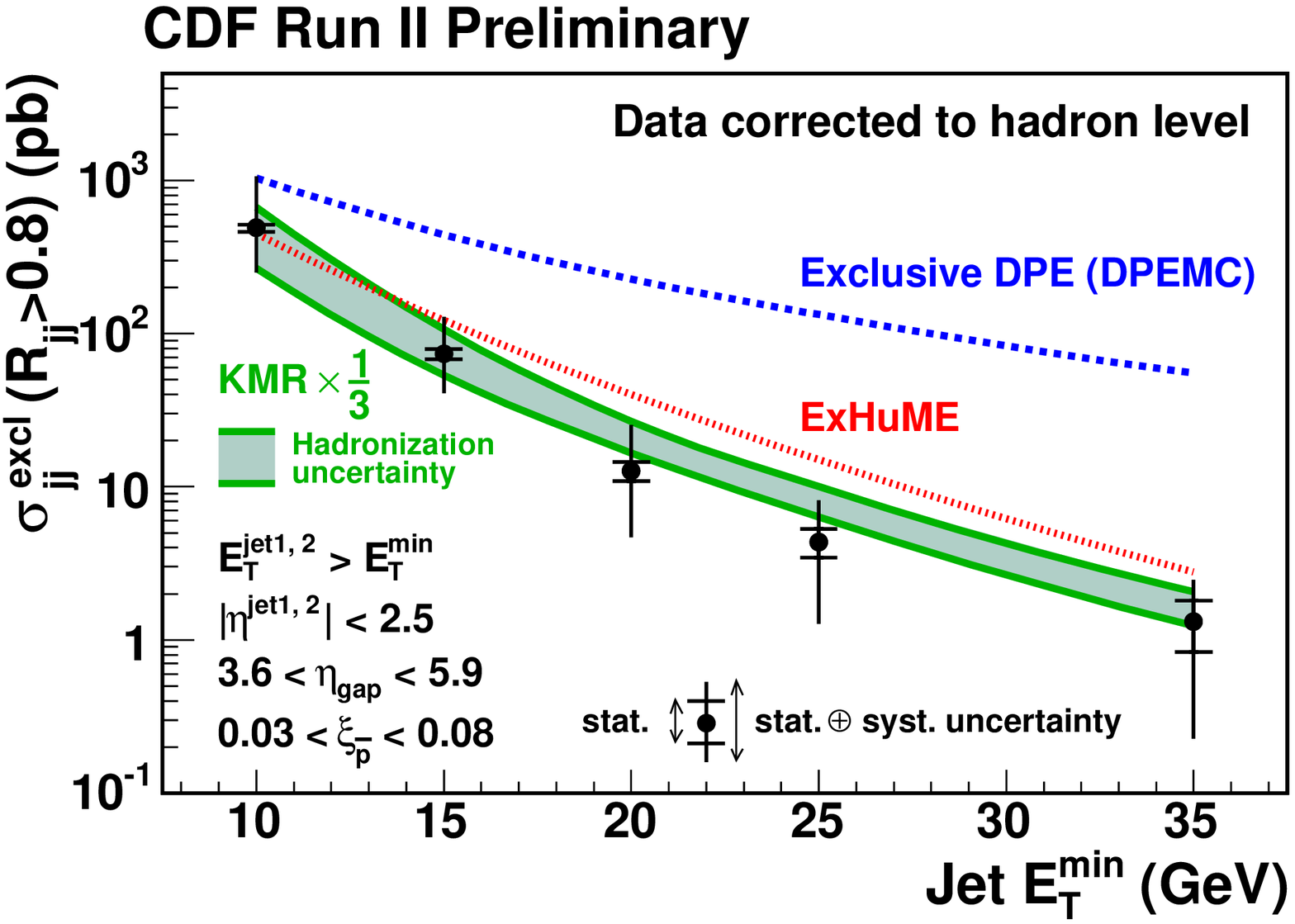}
  %%{plots/BLESS_xsec_excl_etmin_top.eps}
  \caption{(left) Dijet mass fraction $R_{jj}$ in inclusive DPE dijet data
    (left). An excess over predictions at large $R_{jj}$ is observed as a
    signal of exclusive dijet production.
    (right) The measured cross section for exclusive dijet production
    compared to predictions.}
  \label{fig:excl_jj}
\end{figure}

The exclusive dijet production was first searched for by CDF
in Run I data, and the limit of $\sigma_{excl}<3.7$ nb (95\% CL)
was placed~\cite{DPEJJ}.
In the Run II search~\cite{ExclJJ},
first a sample of inclusive DPE dijet events is selected.
The exclusive signal is then searched for examining
dijet mass fraction $R_{jj}$ which is the ratio of dijet mass $M_{jj}$
to system mass $M_X$.
This observable should be sensitive to how much the event energy
is concentrated in the dijet.
The $R_{jj}$ of exclusive dijet events is expected to be peaked around
$R_{jj}\sim 0.8$ and have a long tail toward lower values due to
hadronization effects causing energy leak from jet cones and also the
presence of gluon radiations in the initial and final states.
Figure~\ref{fig:excl_jj} shows $R_{jj}$ distributions for data,
inclusive DPE dijet Monte Carlo (MC) events from {\sc pomwig} Monte
Carlo with various sets of the pomeron structure functions, and the 
non-DPE events.
The data clearly show an excess at high $R_{jj}$ over the
non-DPE background events and inclusive DPE predictions.
The shape of excess is well described by exclusive dijet MC
based on two models (ExHuME~\cite{ExHuME}, DPEMC~\cite{DPEMC});
however, the measured cross section disfavors DPEMC.
Predictions by Khoze {\it et al.}~\cite{KMR} are found to be
consistent with data within its factor of 3 uncertainty.

\section{Summary}

The long-standing diffractive program at CDF has substantially
improved our understanding of diffractive processes.
In Run II, the measurements on diffractive dijets and the diffractive
structure function are extended to $Q^2\sim10^4$ GeV$^2$, and the
measurement of diffractive $W/Z$ production was made using the RP
detector.
The study on events with a rapidity-gap between forward jets is 
underway.
In addition, CDF reported a first observation of exclusive dijet
production which provides a valuable calibration for the predictions
on exclusive Higgs production at the LHC.

%------------------------------------------------------------------------------
%       Bibliography
%------------------------------------------------------------------------------
\begin{footnotesize}
\bibliographystyle{ismd08} 
{\raggedright
\bibliography{ismd08_hard_diffraction}

\providecommand{\etal}{et al.\xspace}
\providecommand{\href}[2]{#2}
\providecommand{\coll}{Coll.}
\catcode`\@=11
\def\@bibitem#1{%
\ifmc@bstsupport
  \mc@iftail{#1}%
    {;\newline\ignorespaces}%
    {\ifmc@first\else.\fi\orig@bibitem{#1}}
  \mc@firstfalse
\else
  \mc@iftail{#1}%
    {\ignorespaces}%
    {\orig@bibitem{#1}}%
\fi}%
\catcode`\@=12
\begin{mcbibliography}{10}

\bibitem{DiffJJ-CDF}
{ CDF} Collaboration, F.~Abe {\em et al.},
\newblock Phys. Rev. Lett.{} {\bf 79},~2636~(1997)\relax
\relax
\bibitem{DiffW-CDF}
{ CDF} Collaboration, F.~Abe {\em et al.},
\newblock Phys. Rev. Lett.{} {\bf 78},~2698~(1997)\relax
\relax
\bibitem{Diffb-CDF}
{ CDF} Collaboration, A.~Affolder {\em et al.},
\newblock Phys. Rev. Lett.{} {\bf 84},~232~(2000)\relax
\relax
\bibitem{RPJJ-CDF}
{ CDF} Collaboration, A.~Affolder {\em et al.},
\newblock Phys. Rev. Lett.{} {\bf 84},~5043~(2000)\relax
\relax
\bibitem{RPJJ-630-CDF}
{ CDF} Collaboration, A.~Affolder {\em et al.},
\newblock Phys. Rev. Lett.{} {\bf 88},~151802~(2002)\relax
\relax
\bibitem{Gotsman:1999xq}
E.~Gotsman, E.~Levin, and U.~Maor,
\newblock Phys. Rev. D{} {\bf 60},~094011~(1999)\relax
\relax
\bibitem{Kaidalov:2001iz}
A.~B. Kaidalov, V.~A. Khoze, A.~D. Martin, and M.~G. Ryskin,
\newblock Eur. Phys. J.{} {\bf C21},~521~(2001)\relax
\relax
\bibitem{Dino}
K.~Goulianos.
\newblock \href{http://www.arXiv.org/abs/hep-ph/0203141}{{\tt
  hep-ph/0203141}}\relax
\relax
\bibitem{DPEJJ}
{ CDF} Collaboration, A.~Affolder {\em et al.},
\newblock Phys. Rev. Lett.{} {\bf 85},~4215~(2000)\relax
\relax
\bibitem{Mike}
M. Albrow, this proceeding\relax
\relax
\bibitem{DiffWZ-D0}
{ D0} Collaboration, V.~Abazov {\em et al.},
\newblock Phys. Lett. B{} {\bf 574},~169~(2003)\relax
\relax
\bibitem{DD-CDF1}
CDF Collaboration, F. Abe {\it et al.}, Phys. Rev. Lett. {\bf 74}, 855
  (1995)\relax
\relax
\bibitem{DD-CDF2}
CDF Collaboration, F. Abe {\it et al.}, Phys. Rev. Lett. {\bf 80}, 1156
  (1998)\relax
\relax
\bibitem{DD-CDF3}
CDF Collaboration, F. Abe {\it et al.}, Phys. Rev. Lett. {\bf 81}, 5278
  (1998)\relax
\relax
\bibitem{DD-CDF4}
{ CDF} Collaboration, A.~Affolder {\em et al.},
\newblock Phys. Rev. Lett.{} {\bf 87},~141802~(2001)\relax
\relax
\bibitem{DD-D0}
{ D0} Collaboration, S.~Abachi {\em et al.},
\newblock Phys. Rev. Lett.{} {\bf 76},~734~(1996)\relax
\relax
\bibitem{Marquet:2007xx}
C.~Marquet and C.~Royon.
\newblock \href{http://www.arXiv.org/abs/arXiv:0704.3409}{{\tt
  arXiv:0704.3409}}\relax
\relax
\bibitem{ExclJJ}
{ CDF} Collaboration, T.~Aaltonen {\em et al.},
\newblock Phys. Rev. D{} {\bf 77},~052004~(2008)\relax
\relax
\bibitem{ExHuME}
J.~Monk and A.~Pilkington,
\newblock Comput. Phys. Commun.{} {\bf 175},~232~(2006)\relax
\relax
\bibitem{DPEMC}
M.~Boonekamp and T.~Kucs,
\newblock Comput. Phys. Commun.{} {\bf 167},~217~(2005)\relax
\relax
\bibitem{KMR}
V.~A. Khoze, A.~D. Martin, and M.~G. Ryskin,
\newblock Eur. Phys. J. C{} {\bf 14},~525~(2000)\relax
\relax
\end{mcbibliography}
}
\end{footnotesize}
\end{document}